\documentclass[aps,twocolumn,superscriptaddress]{revtex4}
\usepackage{graphicx}

\newcommand{\ket}[1] {\left| #1 \right\rangle}

\begin{document}

\title{Superradiance transition in a system with a single qubit and a single oscillator}

\author{S. Ashhab}
\affiliation{Advanced Science Institute, RIKEN, Wako-shi, Saitama
351-0198, Japan}
\affiliation{Physics Department, The University of Michigan, Ann
Arbor, Michigan 48109-1040, USA}

\date{\today}


\begin{abstract}
We consider the phase-transition-like behaviour in the Rabi model
containing a single two-level system, or qubit, and a single
harmonic oscillator. The system experiences a sudden transition
from an uncorrelated state to an increasingly correlated one as
the qubit-oscillator coupling strength is varied and increased
past a critical point. This singular behaviour occurs in the limit
where the oscillator's frequency is much lower than the qubit's
frequency; away from this limit one obtains a finite-width
transition region. By analyzing the energy-level structure, the
value of the oscillator field and its squeezing and the
qubit-oscillator correlation, we gain insight into the nature of
the transition and the associated critical behaviour.
\end{abstract}


\maketitle

\section{Introduction}

The interaction between light and matter, and more generally
between harmonic oscillators and few-level quantum systems, is
ubiquitous in nature. In spite of the simplicity in its basic
mathematical description, it results in a wide variety of
phenomena, some of which have been analyzed in detail over the
past few decades \cite{QuantumOpticsBooks}.

One of the interesting phenomena involving light-matter
interaction is superradiance. The study of this phenomenon started
with the idea that an ensemble containing a large number of atoms
can exhibit quantum-coherent collective behaviour in its
absorption and emission of photons \cite{Dicke}. This observation
gave rise to the Dicke model, where a large number of atoms
interact with a single (harmonic-oscillator) mode of the
electromagnetic field. It was later realized that the Dicke model
exhibits a phase transition, both thermal and quantum, between a
state with negligible light-matter correlations and one with
strong correlations \cite{Hepp,Wang,Emary,Sachdev}. In the case of
the quantum phase transition, the correlations appear when the
coupling strength between the two subsystems exceeds a certain
critical value.

The prediction of the superradiance phase transition in the Dicke
model has resulted in enormous interest, including a debate that
continues to this day \cite{Rzazewski} on whether such a phase
transition could occur for a system with the usual electric
coupling between light and matter.

Studies on the superradiance phase transition in the Dicke model
typically consider the thermodynamic limit, where the number of
atoms approaches infinity with the effective coupling strength
between the collective atomic mode and the electromagnetic mode
kept independent of atom number. This approach to realizing the
phase transition is naturally motivated by the fact that the
interaction between natural atoms and optical-frequency cavities
is weak compared to the bare atomic and cavity frequencies.

Recently, the realization of qubit-oscillator systems using
superconducting qubit circuits has made it possible to achieve the
so-called ultrastrong-coupling regime, where the coupling strength
between a single qubit and a single oscillator is comparable to
the bare frequencies of the two constituents
\cite{Niemczyk,FornDiaz,Devoret,SCQubitReviews,OtherSystems}. This
ability relaxes the requirement of using large atomic ensembles in
order to study strong-coupling effects; a single (artificial) atom
suffices. Since the artificial atom in these studies is
effectively a two-level system, we shall sometimes refer to it as
the qubit.

It has been noted in a number of recent studies that the
single-qubit--single-oscillator problem exhibits similar
phase-transition-like behaviour in the limit where the ratio of
the resonator frequency to the qubit frequency tends to zero
\cite{Levine,Ashhab,Hwang,Bakemeier}. Here we examine this limit
closely and analyze the associated transition. We do so by
analyzing the behaviour of several physical quantities in the
transition region. These include the energy-level structure, the
average value of the field in the cavity, the squeezing in the
oscillator and the qubit-oscillator entanglement. It should be
emphasized that the limit considered here is clearly distinct from
the thermodynamic limit with a large number of qubits. We shall
therefore not use the term ``phase transition" in this paper. It
is quite interesting that several ground-state properties exhibit
essentially the same behaviour in the two distinct limits. The
correspondence between the two limits is not complete, however, as
evidenced by the fact that no thermal phase transition occurs in
the simple system considered here (see Sec.~V).

\section{Model Hamiltonian}

The system that we consider here is composed of a single qubit
coupled to a single harmonic oscillator. The coupling contains
only one term, and this term is linear in the oscillator
variables. The Hamiltonian describing this quantum system is given
by
\begin{equation}
\hat{H} = \hat{H}_{\rm q} + \hat{H}_{\rm ho} + \hat{H}_{\rm int},
\label{Eq:Basic_Hamiltonian}
\end{equation}
where
\begin{eqnarray}
\hat{H}_{\rm q} & = & -\frac{\Delta}{2} \hat{\sigma}_x -
\frac{\epsilon}{2} \hat{\sigma}_z \nonumber
\\
\hat{H}_{\rm ho} & = & \hbar \omega_0 \hat{a}^{\dagger} \hat{a} +
\frac{1}{2} \hbar \omega_0
\label{Eq:Detailed_Hamiltonians}
\\
\hat{H}_{\rm int} & = & \lambda \left( \hat{a} + \hat{a}^{\dagger}
\right) \hat{\sigma}_z, \nonumber
\end{eqnarray}
$\hat{\sigma}_{x}$ and $\hat{\sigma}_{z}$ are the usual Pauli
matrices (with $\hat{\sigma}_z\ket{\uparrow}=\ket{\uparrow},
\hat{\sigma}_z\ket{\downarrow}=-\ket{\downarrow}$), and
$\hat{a}^{\dagger}$ and $\hat{a}$ are the creation and
annihilation operators of the harmonic oscillator. The parameters
$\Delta$ and $\epsilon$ are the so-called gap and bias of the
qubit, $\omega_0$ is the oscillator's characteristic frequency,
and $\lambda$ is the qubit-oscillator coupling strength.

For purposes of the present study, we focus on the case with the
qubit biased at its symmetry point ($\epsilon=0$), where the
Hamiltonian can simply be expressed as
\begin{equation}
\hat{H} = -\frac{\Delta}{2} \hat{\sigma}_x + \hbar \omega_0
\hat{a}^{\dagger} \hat{a} + \lambda \left( \hat{a} +
\hat{a}^{\dagger} \right) \hat{\sigma}_z.
\label{Eq:Basic_Hamiltonian_SymmetryPoint}
\end{equation}
It should also be noted that in writing this Hamiltonian [and also
in Eq.~(\ref{Eq:Detailed_Hamiltonians})] we ignore the so-called
$A^2$ term that is at the heart of the
superradiance-phase-transition controversy \cite{Rzazewski}.

In the absence of coupling, i.e.~when $\lambda=0$, the ground
state of the system is given by
\begin{equation}
\ket{\rm GS}_{\lambda=0} =
\frac{\ket{\uparrow}+\ket{\downarrow}}{\sqrt{2}} \otimes \ket{\rm
vac}.
\label{Eq:Ground_state_zero_coupling}
\end{equation}
For very strong coupling, the quantitative definition of which
will become clear below, the ground state is highly correlated and
(to a good approximation) given by
\begin{equation}
\ket{\rm GS}_{({\rm large \ }\lambda)} = \frac{1}{\sqrt{2}} \left(
\ket{\uparrow} \otimes \ket{-\alpha} + \ket{\downarrow} \otimes
\ket{\alpha} \right),
\label{Eq:Ground_state_infinite_coupling}
\end{equation}
where $\ket{\pm\alpha}$ are coherent states with opposite values
of the oscillator variable that couples to the qubit, i.e.~the
field operator $(\hat{a}+\hat{a}^{\dagger})/2$.

\section{Transition point and critical behaviour}

Using the results from the thermodynamic limit, i.e.~the limit
where the number of atoms is large, the critical coupling strength
separating the uncorrelated and correlated ground states is
expected to occur at the point
\begin{equation}
\lambda_{\rm c} = \frac{\sqrt{\hbar\omega_0\Delta}}{2}.
\label{Eq:CriticalCouplingStrength}
\end{equation}
We shall show through the behaviour of various quantities that
similar behaviour is obtained in the single-atom case in the limit
$\hbar\omega_0/\Delta\rightarrow 0$. It should be emphasized that,
even though one might worry about the possibility that
$\lambda_{\rm c}$ might vanish or diverge in this limit, this
apparent complication disappears if one treats $\lambda_{\rm c}$
as a reference point for measuring the coupling strength. If one
then considers the behaviour of the system as the parameter
$\lambda/\lambda_{\rm c}$ is varied across the point
$\lambda/\lambda_{\rm c}=1$, no complications related to the
behaviour of $\lambda_{\rm c}$ arise.

\begin{figure}[h]
\includegraphics[width=9.0cm]{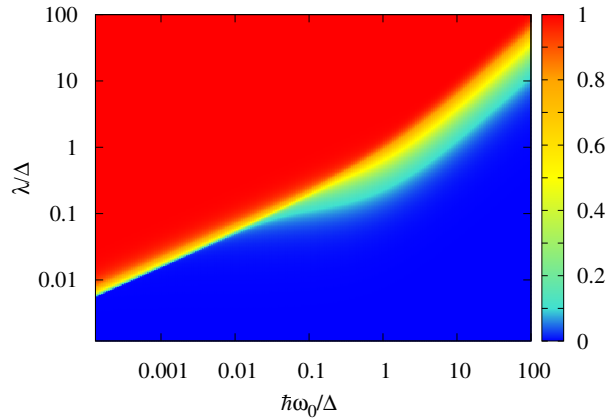}
\caption{(Color online) The von Neumann entropy $S$ as a function
of the oscillator frequency $\hbar\omega_0$ and the coupling
strength $\lambda$, both measured in comparison to the qubit
frequency $\Delta$. One can see clearly that moving in the
vertical direction the rise in entropy is sharp in the regime
$\hbar\omega_0/\Delta\ll 1$, whereas it is smooth when
$\hbar\omega_0/\Delta$ is comparable to or larger than 0.1.}
\label{Fig:ColorPlot}
\end{figure}

\begin{figure}[h]
\includegraphics[width=7.5cm]{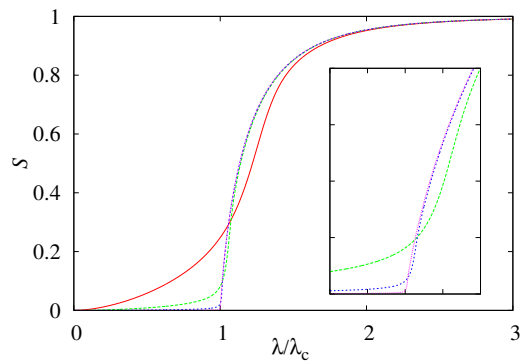}
\caption{(Color online) The von Neumann entropy $S$ as a function
of coupling strength $\lambda$ [measured in comparison to the
critical coupling strength given by
Eq.~(\ref{Eq:CriticalCouplingStrength})] for various values of the
ratio $\hbar\omega_0/\Delta$. In particular, we use the values
$10^{-1}$ (red solid line), $10^{-2}$ (green dashed line),
$10^{-3}$ (blue short-dashed line) and $10^{-4}$ (purple dotted
line). The inset shows a magnified plot in the region around the
critical point: The ranges of the $x$ and $y$ axes are [0.9,1.1]
and [0,0.4], respectively. As the ratio $\hbar\omega_0/\Delta$
decreases, the onset of entropy becomes increasingly sudden.}
\label{Fig:EntropyLinear}
\end{figure}

\begin{figure}[h]
\includegraphics[width=7.5cm]{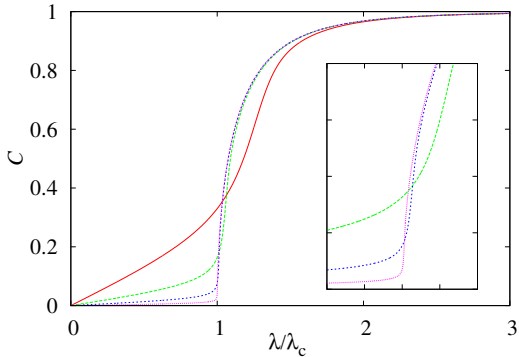}
\caption{(Color online) The correlation function $C=\langle
\sigma_z {\rm sign}(a+a^{\dagger})\rangle$ as a function of
coupling strength $\lambda$ [measured in comparison to the
critical coupling strength given by
Eq.~(\ref{Eq:CriticalCouplingStrength})] for various values of the
ratio $\hbar\omega_0/\Delta$: $10^{-1}$ (red solid line),
$10^{-2}$ (green dashed line), $10^{-3}$ (blue short-dashed line)
and $10^{-4}$ (purple dotted line). The inset shows a magnified
plot in the region around the critical point: The ranges of the
$x$ and $y$ axes are [0.9,1.1] and [0,0.4], respectively. The
correlation function $C$ exhibits behaviour similar to that of the
entropy, which is shown in Fig.~\ref{Fig:EntropyLinear}.}
\label{Fig:SpinFieldSignCorrelationFunction}
\end{figure}

\begin{figure}[h]
\includegraphics[width=6.5cm]{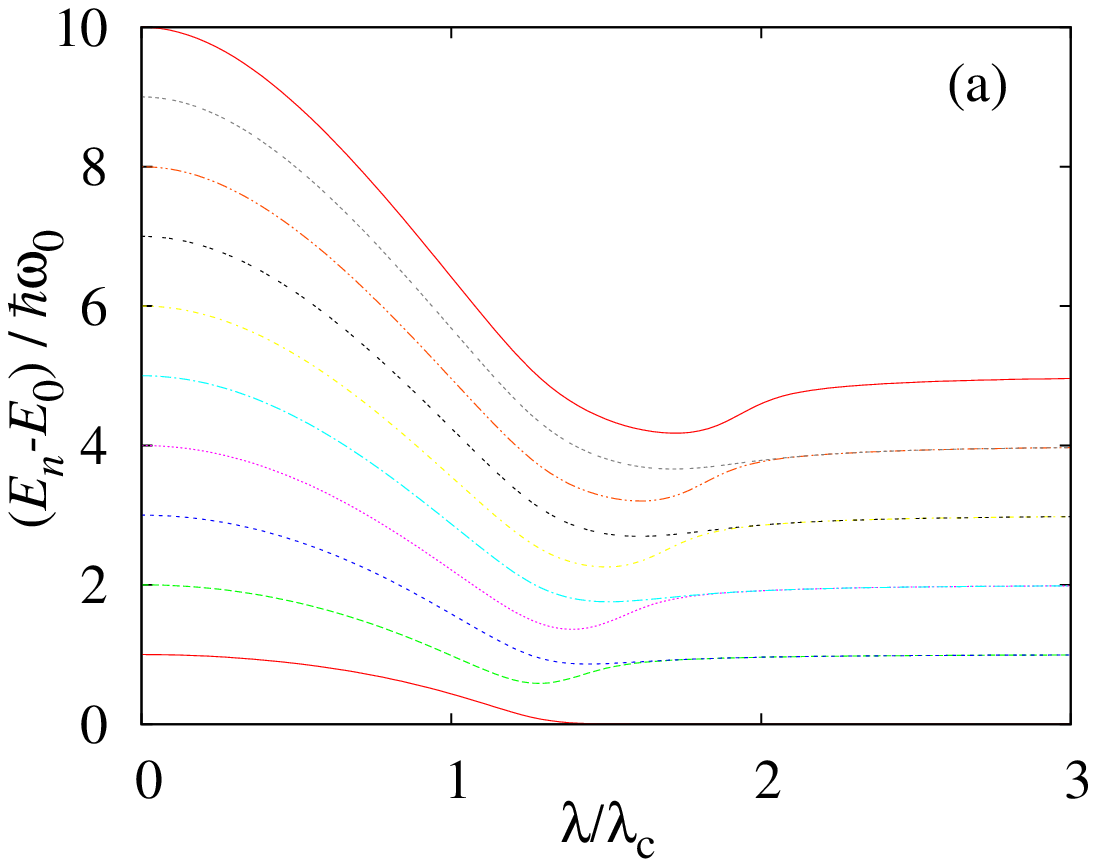}
\includegraphics[width=6.5cm]{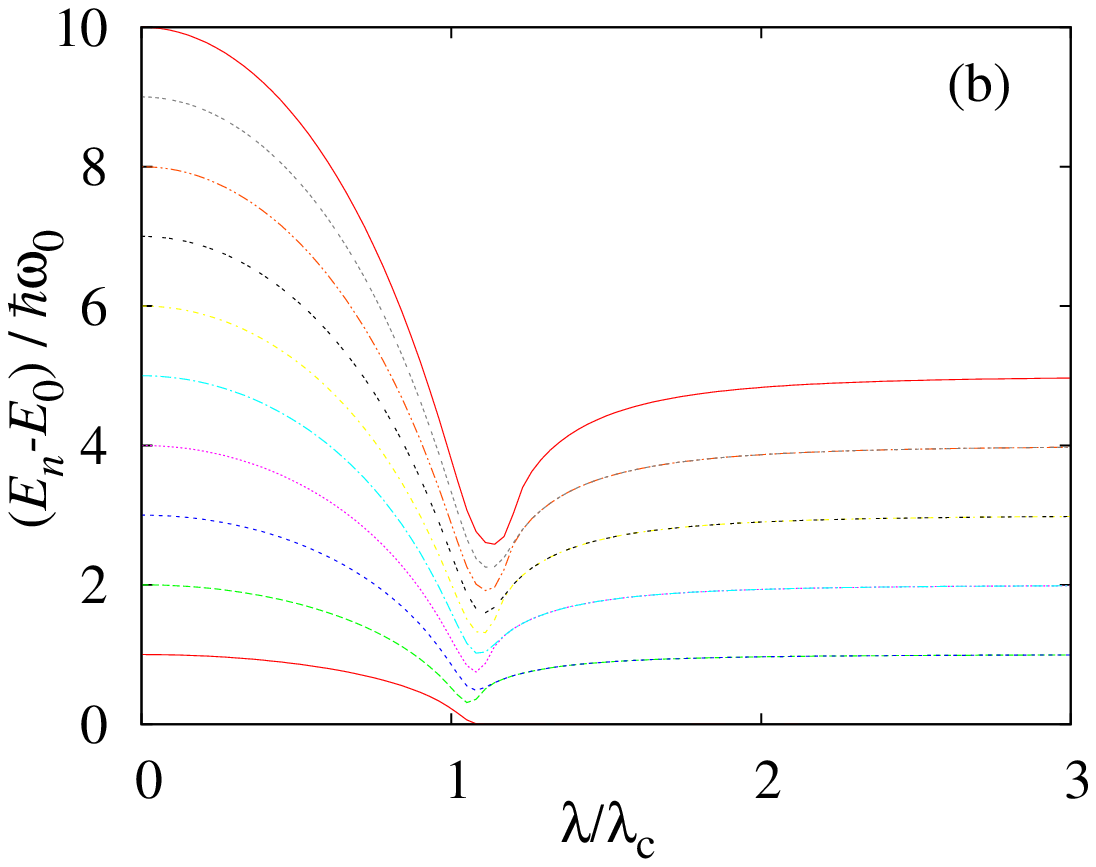}
\includegraphics[width=6.5cm]{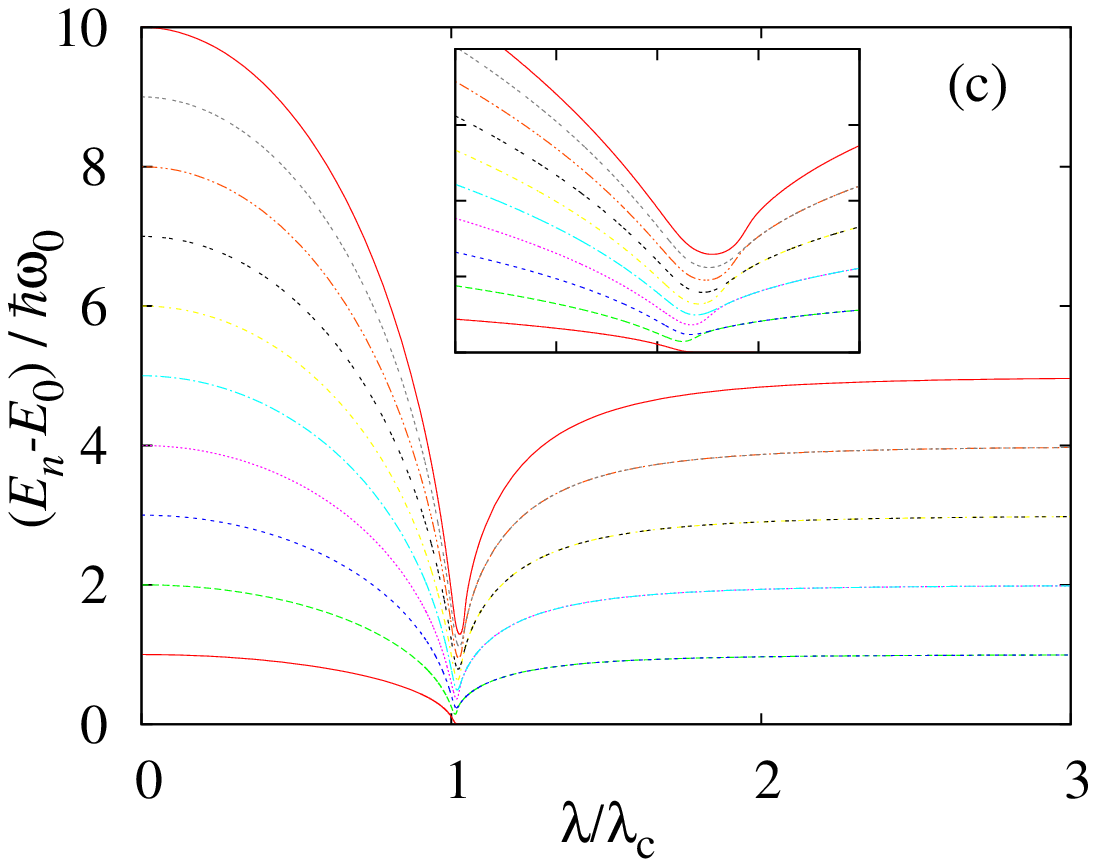}
\caption{(Color online) The energy levels of the first ten excited
states relative to the ground state energy as functions of the
coupling strength $\lambda$ [measured in comparison to the
critical coupling strength given by
Eq.~(\ref{Eq:CriticalCouplingStrength})]. In (a) we take
$\hbar\omega_0/\Delta=10^{-1}$, in (b) we take
$\hbar\omega_0/\Delta=10^{-2}$, and in (c) we take
$\hbar\omega_0/\Delta=10^{-3}$. The energy levels become
increasingly dense as the coupling strength approaches the
critical value, and they separate again (forming pairs) after the
coupling strength exceeds the critical value. The inset in (c)
shows a magnified plot in the region around the critical point:
the ranges of the x and y axes are [0.9,1.1] and [0,4],
respectively.}
\label{Fig:EnergyLevels}
\end{figure}

\begin{figure}[h]
\includegraphics[width=6.5cm]{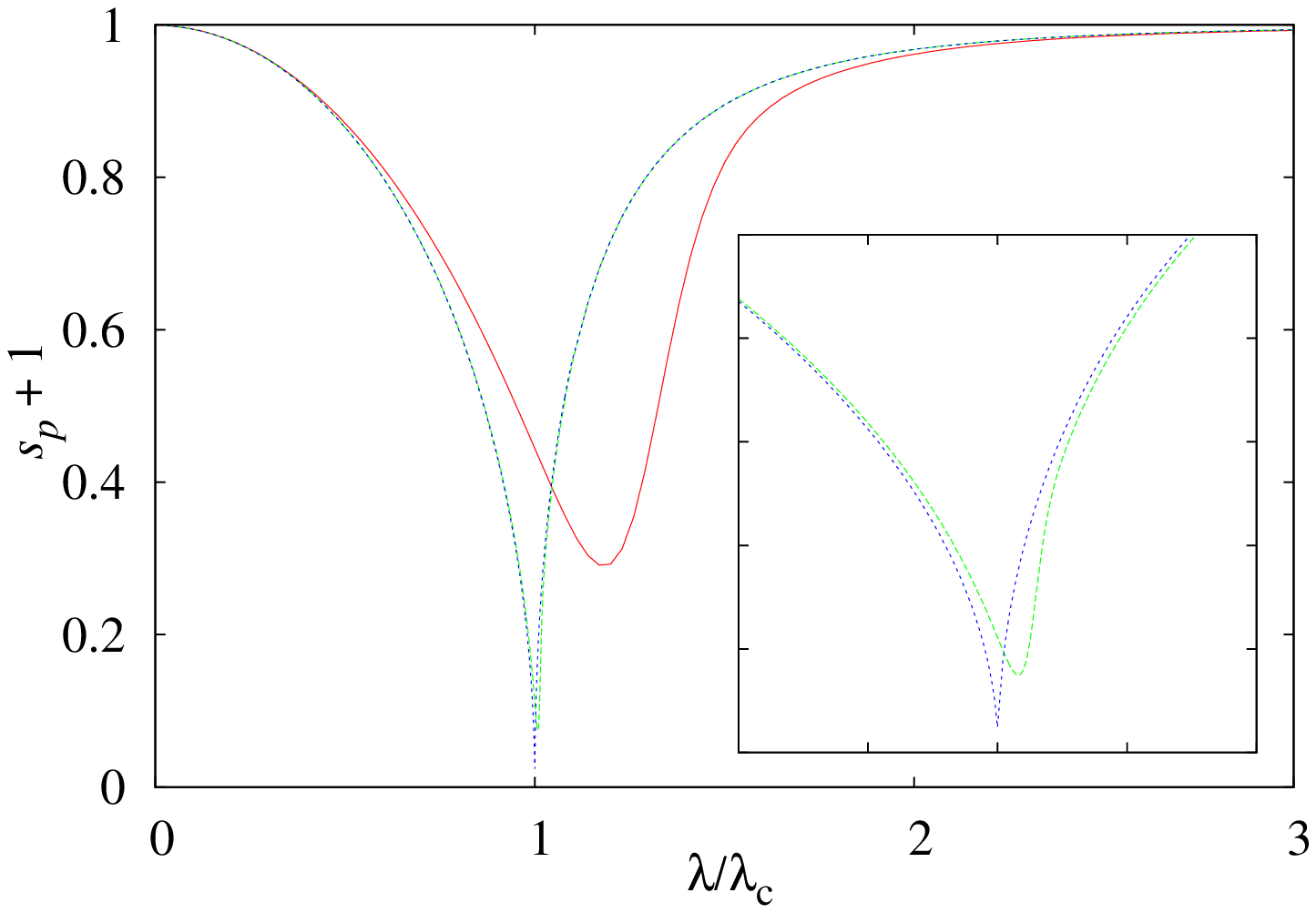}
\caption{(Color online) The squeezing quantifier $s_p+1$ as a
function of the coupling strength $\lambda$ [measured in
comparison to the critical coupling strength given by
Eq.~(\ref{Eq:CriticalCouplingStrength})]. The different lines
correspond to $\hbar\omega_0/\Delta=10^{-1}$ (solid red line),
$10^{-3}$ (dashed green line) and $10^{-5}$ (short-dashed blue
line). The inset shows a magnified plot in the region around the
critical point: the ranges of the x and y axes are [0.9,1.1] and
[0,0.5], respectively. The behaviour of the squeezing parameter
mirrors that of the low-lying excited states shown in
Fig.~\ref{Fig:EnergyLevels}.}
\label{Fig:SqueezingParameter}
\end{figure}

The tendency towards singular behaviour (in the dependence of
various physical quantities on $\lambda$) in the limit
$\hbar\omega_0/\Delta\rightarrow 0$ is illustrated in
Figs.~\ref{Fig:ColorPlot}-\ref{Fig:SqueezingParameter}. In these
figures, the entanglement, spin-field correlation function,
low-lying energy levels (measured from the ground state) and the
oscillator's squeezing parameter are plotted as functions of the
coupling strength. It is clear from Figs.~\ref{Fig:EntropyLinear}
and \ref{Fig:SpinFieldSignCorrelationFunction} that when
$\hbar\omega_0/\Delta\leq 10^{-3}$ both the entanglement (which is
quantified through the von Neumann entropy $S={\rm Tr}\{\rho_{\rm
q}\log_2\rho_{\rm q}\}$ with $\rho_{\rm q}$ being the qubit's
reduced density matrix) and the correlation function $C=\langle
\sigma_z {\rm sign}(a+a^{\dagger})\rangle$ rise sharply upon
crossing the critical point
\cite{EntropyVersusCorrelationFunctionFootnote}. The low-lying
energy levels, shown in Fig.~\ref{Fig:EnergyLevels}, approach each
other to form a large group of almost degenerate energy levels at
the critical point before they separate again into pairs of
asymptotically degenerate energy levels. This approach is not
complete, however, even when $\hbar\omega_0/\Delta=10^{-3}$; for
this value the energy level spacing in the closest-approach region
is roughly ten times smaller than the energy level spacing at
$\lambda=0$. The squeezing parameter is defined by the width of
the momentum distribution relative to that in the case of an
isolated oscillator. For consistency with Ref.~\cite{Ashhab}, we
define it as
\begin{equation}
s_p + 1 = \frac{\langle \hat{p}^2 \rangle}{\left. \langle
\hat{p}^2 \rangle \right|_{\lambda=0}},
\end{equation}
where $\hat{p}$ is the oscillator's momentum operator, which is
proportional to $i(\hat{a}^{\dagger}-\hat{a})$ in our definition
of the operators. The squeezing parameter mirrors the behaviour of
the low-lying energy levels. In particular we can see from
Fig.~\ref{Fig:SqueezingParameter} that only when
$\hbar\omega_0/\Delta$ reaches the value $10^{-5}$ does the
squeezing become almost singular at the critical point.

\begin{figure}[h]
\includegraphics[width=7.0cm]{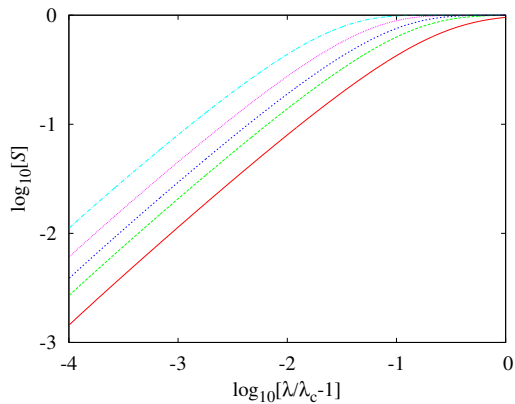}
\caption{(Color online) The logarithm of the von Neumann entropy
$S$ as a function of the logarithm of the quantity
$(\lambda/\lambda_{\rm c})-1$, which measures the distance of the
coupling strength from the critical value. The red solid line
corresponds to the single-qubit case, whereas the other lines
correspond to the multi-qubit case: $N=2$ (green dashed line), 3
(blue short-dashed line), 5 (purple dotted line) and 10
(dash-dotted cyan line). All the lines correspond to
$\hbar\omega_0/\Delta=10^{-7}$. The slope of all lines is
approximately 0.92 when $(\lambda/\lambda_{\rm c})-1=10^{-4}$. The
ratio of the entropy in the multi-qubit case to that in the
single-qubit case approaches $N$ for all the lines as we approach
the critical point.}
\label{Fig:EntropyLogLog}
\end{figure}

\begin{figure}[h]
\includegraphics[width=7.0cm]{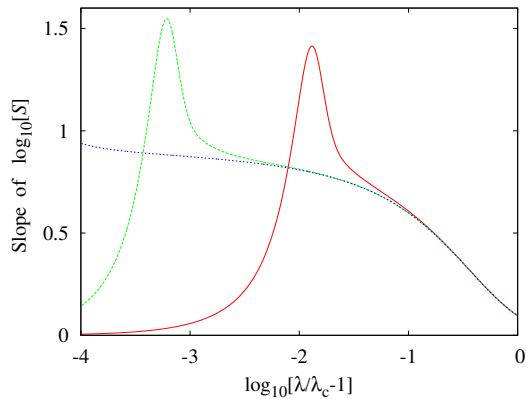}
\caption{(Color online) The slope of the logarithm of the von
Neumann entropy $S$ as a function of the logarithm of the quantity
$(\lambda/\lambda_{\rm c})-1$, which measures the distance of the
coupling strength from the critical value. The different lines
correspond to $\hbar\omega_0/\Delta=10^{-3}$ (solid red line),
$10^{-5}$ (dashed green line) and $10^{-7}$ (short-dashed blue
line). The slope seems to be approaching the value 1 as we
approach the critical point, but at some point, determined by the
ratio $\hbar\omega_0/\Delta$, the slope has a peak and drops to
zero. The deviation from the simple asymptotic behaviour is
related to the fact that $S$ has a non-zero value at
$\lambda/\lambda_c=1$, as can be seen in
Fig.~\ref{Fig:EntropyLinear}.}
\label{Fig:EntropyLogLogSlope}
\end{figure}

We now look more closely at the critical exponents around the
critical point. Analytical expressions describing the critical
behaviour of some quantities can be obtained using a semiclassical
calculation \cite{Ashhab}. In particular, this approximation gives
the result that just below the critical point, the energy-level
separation has the functional dependence
\begin{equation}
E_n-E_{n-1} = \sqrt{2} \hbar\omega_0 \left( 1 -
\frac{\lambda}{\lambda_{\rm c}} \right)^{1/2},
\end{equation}
while just above the critical point it is given by
\begin{equation}
E_n-E_{n-2} = 2 \hbar\omega_0 \left( \frac{\lambda}{\lambda_{\rm
c}} - 1 \right)^{1/2}.
\end{equation}
The squeezing parameter should exhibit the same behaviour. The
numerical results shown in Figs.~\ref{Fig:EnergyLevels} and
\ref{Fig:SqueezingParameter} approach this functional dependence
on both sides of the critical point as we decrease the ratio
$\hbar\omega_0/\Delta$. Under the same semiclassical
approximation, one can also analytically calculate the average
value of the field, i.e.~$\langle (\hat{a}+\hat{a}^{\dagger})/2
\rangle$ [In the superradiance region one calculates the value in
one of the two branches of the wave function, i.e.~one calculates
the value of $\alpha$ in
Eq.~(\ref{Eq:Ground_state_infinite_coupling})]. The average value
of the field vanishes below the critical point, and it has the
form
\begin{equation}
\alpha = \frac{\Delta}{4\lambda} \left[
\left(\frac{\lambda}{\lambda_{\rm c}}\right)^4 - 1 \right]^{1/2}
\end{equation}
above the critical point. Its dependence on the coupling strength
just above the critical point can alternatively be expressed as
\begin{equation}
\alpha = \frac{\Delta}{2\lambda_{\rm c}} \left(
\frac{\lambda}{\lambda_{\rm c}} - 1 \right)^{1/2}.
\end{equation}
In Fig.~\ref{Fig:EntropyLogLog} we plot the von Neumann entropy as
a function of coupling strength on a log-log scale (above the
critical point). The slope of the curve for the smallest values of
$\lambda$ is approximately 0.92. This value suggests that the true
critical exponent might be unity. One difficulty in calculating
the asymptotic value of the slope is the fact that for any finite
value of $\hbar\omega_0/\Delta$, the von Neumann entropy deviates
from the behaviour shown in Fig.~\ref{Fig:EntropyLogLog} if one
comes too close to the critical point. This deviation can be seen
in Fig.~\ref{Fig:EntropyLinear} and is illustrated more clearly in
Fig.~\ref{Fig:EntropyLogLogSlope}.

\section{Multi-qubit case}

Let us now consider the case with a finite number $N$ of qubits
\cite{FiniteSizeEffects}. In this model, the qubits are usually
assumed to have the same single-qubit energies $\Delta$ and the
same coupling strength to the oscillator, which is usually defined
as $\lambda/\sqrt{N}$. The Hamiltonian in this case is given by
\begin{eqnarray}
\hat{H} & = & -\sum_{j=1}^{N} \frac{\Delta}{2}
\hat{\sigma}_x^{(j)} + \hbar \omega_0 \hat{a}^{\dagger} \hat{a} +
\sum_{j=1}^{N} \frac{\lambda}{\sqrt{N}} \left( \hat{a} +
\hat{a}^{\dagger} \right) \hat{\sigma}_z^{(j)}
\nonumber \\
& = & - \Delta \hat{J}_x + \hbar \omega_0 \hat{a}^{\dagger}
\hat{a} + 2 \frac{\lambda}{\sqrt{N}} \left( \hat{a} +
\hat{a}^{\dagger} \right) \hat{J}_z,
\label{Eq:N-qubit_Hamiltonian}
\end{eqnarray}
where we have defined the total spin operators
$\hat{J}_{\alpha}=\sum\hat{\bf \sigma}_{\alpha}/2$. In the limit
$\hbar\omega_0/\Delta\rightarrow 0$, all the results concerning
the low-energy spectrum of the resonator remain unchanged; one
could say that the reduction of the coupling strength by the
factor $\sqrt{N}$ is compensated by the strengthening of the spin
raising and lowering operators by the same factor because of the
collective behaviour of the qubits. In particular, the transition
occurs at the critical coupling strength given by
Eq.~(\ref{Eq:CriticalCouplingStrength}). Because the qubits now
have a larger total spin (when compared to the single-qubit case),
spin states that are separated by small angles can be drastically
different (i.e.~have a small overlap). In particular, the overlap
for $N$ qubits is given by $\cos^{2N}(\theta/2)$. By expanding
this function to second order around $\theta=0$, one can see that
for small values of $\theta$ the relevant overlap is lower than
unity by an amount that is proportional to $N$. This dependence
translates into the dependence of the qubit-oscillator
entanglement on the coupling strength just above the critical
point. The entanglement therefore rises more sharply in the
multi-qubit case (with the increase being by a factor $N$), as
demonstrated in Fig.~\ref{Fig:EntropyLogLog}.

\section{Finite-temperature behaviour}

Equation (\ref{Eq:Ground_state_infinite_coupling}) gives the
ground state deep in the superradiance region. The first-excited
state has the same form, but with a minus sign instead of the plus
sign. The energy separation between these two states decreases
exponentially with increasing $\lambda$. As a result, an
infinitesimally small temperature would be sufficient to destroy
the coherence between the two branches of the wave function in
thermodynamic equilibrium. Nevertheless, the correlation function
$C$ exhibits essentially the same behaviour for the two states.
Furthermore, all low-lying energy levels have a qualitatively
similar correlation between the state of the qubit and the field
in the oscillator (even though the entanglement might be lost).
One can therefore ask whether a finite-temperature phase
transition would still occur between a region of correlated
qubit-oscillator states and a region with no correlation.

The energy level structure in the single-qubit case is simple in
principle. In the limit $\hbar\omega_0/\Delta\rightarrow 0$, one
can say that the energy levels form two sets, one corresponding to
each qubit state. Each one of these sets has a structure that is
similar to that of a harmonic oscillator with some modifications
that are not central in the present context. In particular the
density of states has a weak dependence on energy, a situation
that cannot support a thermal phase transition. If the temperature
is increased while all other system parameters are kept fixed,
qubit-oscillator correlations (which are finite only above the
critical point) gradually decrease and vanish asymptotically in
the high-temperature limit. No singular point is encountered along
the way. This result implies that the transition point is
independent of temperature. In other words, it remains at the
value given by Eq.~(\ref{Eq:CriticalCouplingStrength}) for all
temperatures. If, for example, one is investigating the dependence
of the correlation function $C$ on the coupling strength (as
plotted in Fig.~\ref{Fig:SpinFieldSignCorrelationFunction}), the
only change that occurs as we increase the temperature is that the
qubit-oscillator correlations change more slowly when the coupling
strength is varied.

\section{Relation to phase transition in the thermodynamic limit}

As we have mentioned above, the singular transition in the limit
$\hbar\omega_0/\Delta\rightarrow 0$ is distinct from that
encountered in the thermodynamic limit $N\rightarrow\infty$. Given
the similarities between the two transitions, one can ask whether
it is possible to identify a single, unified condition for the
realization of singular behaviour. For example, one candidate for
this unified condition could be
$\hbar\omega_0/(N\Delta)\rightarrow 0$.

If such a unified condition existed, we would expect that for any
large value of $N$ there is a proportionately large value of
$\hbar\omega_0/\Delta$ above which the sharp transition is
replaced by a smooth crossover. However, the presence of a phase
transition in the thermodynamic limit is independent of the ratio
$\hbar\omega_0/\Delta$, including the zero and infinite limits. In
particular, if we consider the Dicke model with an arbitrary value
of $N$ and first take the limit
$\hbar\omega_0/\Delta\rightarrow\infty$ (meaning that this is the
most dominant infinite limit in the problem), the system
effectively reduces to the Lipkin-Meshkov-Glick model, which
exhibits singular behaviour in the limit $N\rightarrow\infty$ (see
e.g.~Refs.~\cite{Bakemeier,Tsomokos}). We therefore conclude that
the two limits $\hbar\omega_0/\Delta\rightarrow 0$ and
$N\rightarrow\infty$ cannot be unified in a nontrivial manner.

In fact, the consideration of the thermodynamic limit provides at
least a partial explanaton for why singular behaviour is obtained
only in the limit $\hbar\omega_0/\Delta\rightarrow 0$ in the
single-qubit case. In the limit $N\rightarrow\infty$, the phase
transition occurs independently of the ratio
$\hbar\omega_0/\Delta$. However, depending on the value of this
ratio, the phase transition region involves larger changes in the
lower-frequency subsystem, i.e.~either the collective state of the
qubits or the state of the oscillator. In the limit
$\hbar\omega_0/\Delta\rightarrow 0$, the state of the qubits only
slightly deviates from the ground state when the transition point
is crossed, and it is plausible that the singular behaviour would
persist even when the ensemble of qubits is replaced by a single
qubit. In the limit $\hbar\omega_0/\Delta\rightarrow\infty$, the
oscillator stays close to its ground state while the state of the
qubit ensemble undergoes large changes upon crossing the
transition point, a behaviour that clearly cannot translate
straightforwardly to the single-qubit case.

\section{Conclusion}

We have analyzed the transition from an uncorrelated composite
system to superradiance behaviour in the
single-qubit--single-oscillator Rabi model. We have shown that as
the ratio of the oscillator's frequency to the qubit's frequency
approaches zero, various physical quantities exhibit singular
dependence that closely resembles that encountered in the study of
the superradiance phase transition in the thermodynamic limit of
the Dicke model.

The qubit-oscillator entanglement and appropriate qubit-oscillator
correlation functions remain very small below the transition point
but increase rapidly as soon as the coupling strength exceeds a
certain critical value. The low-lying energy levels (almost)
collapse to a single highly degenerate ground-state manifold at
the transition point. The amount of squeezing also peaks in a
singular manner at the critical point.

The energy-level separations and the degree of squeezing scale as
$|\lambda/\lambda_c-1|^{1/2}$ on both sides of the critical point,
while the qubit-oscillator entanglement rises as
$|\lambda/\lambda_c-1|^{\alpha}$ above the critical point, with
the exponent $\alpha$ being slightly below unity.

In spite of the similarities in the behaviour of this system with
the behaviour of the Dicke model in the thermodynamic limit, the
analogy is not complete, as evidenced by the absence of a thermal
phase transition in the single-qubit--single-oscillator system.

The Rabi model with arbitrary coupling strength remains an active
area of research. Recent studies have addressed questions related
to the integrability of the model \cite{Braak}, various
approximations and exact solutions
\cite{Hwang,ApproximationsForRabiModel}, dynamics and dissipation
\cite{DynamicsAndDissipationStudies}, proposals of potentially
robust designs for quantum bits \cite{Nataf} and novel strongly
correlated many-polariton states \cite{Schiro}. The present work
deals with this ubiquitous physical model, and we expect that the
results presented here will help improve our understanding of the
basic properties of the model.

\section*{Acknowledgments}

We would like to thank P. Forn-D\'iaz, J. R. Johansson and N. W.
Lambert for useful discussions. This work was supported in part by
ARO, NSF Grant No.~0726909, JSPS-RFBR Contract No.~09-02-92114,
Grant-in-Aid for Scientific Research (S), MEXT Kakenhi on Quantum
Cybernetics, and the JSPS via its FIRST program.


\begin{thebibliography}{99}
\bibitem{EntropyVersusCorrelationFunctionFootnote} The entropy $S$ and the
correlation function $C$ exhibit qualitatively similar features.
Even though (for the same value of $\hbar\omega_0/\Delta$) the
features are less pronounced in the case of $C$ as compared to
$S$, they are more robust. In particular, even if one considers a
case where the first-excited state becomes substantially
populated, the entanglement can be drastically reduced [since for
example an equal mixture of the ground and first-excited state is
equivalent to a mixture of the two terms in
Eq.~(\ref{Eq:Ground_state_infinite_coupling})], while the
correlation function $C$ is only slightly modified. In other
words, an experiment might fail to observe the results plotted in
Fig.~\ref{Fig:EntropyLinear} but still observe those plotted in
Fig.~\ref{Fig:SpinFieldSignCorrelationFunction} (up to some small
imperfections). The reason behind this robustness is the fact that
all low-lying energy eigenstates possess similar correlations
between the spin and the cavity field.
\end{thebibliography}
\end{document}